\documentclass[12 pt]{article}

\usepackage{lmodern}
\usepackage{amsfonts}
\usepackage{amsmath}
\usepackage{graphicx}
\usepackage{subfigure}
\usepackage{wasysym}
\usepackage{titlesec}
\usepackage{float}
\usepackage[margin=1in]{geometry}
\usepackage{pdfpages}
\usepackage{sectsty}
\usepackage[font=large,labelfont=bf]{caption}

\sectionfont{\fontsize{12}{16}\selectfont}

\renewenvironment{abstract}
{\par\noindent\textbf{\abstractname \\}\ \ignorespaces}
{\par\medskip}
\let\tit\textit
\let\tbf\textbf

\begin{document}

\title{\Large \textbf{Generalized Fractal Dimension for a Dissipative Multi-fractal Cascade Model for Fully Developed Turbulence }}
\author{\normalsize Bhimsen Shivamoggi, Michael Undieme, Zoe Barbeau, and Angela Colbert} 

\date{\normalsize University of Central Florida \\ Orlando, Florida 32816  }

\maketitle

\noindent 
\begin{abstract}
\normalsize
\noindent In this paper (Shivamoggi et al. \cite{shivamoggi13}), we explore a variant for the simple model based on a binomial multiplicative process of Meneveau and Sreenivasan \cite{ms9} that mimics the multi-fractal nature of the energy dissipation field in the inertial range of fully developed turbulence (FDT), and uses the generalized fractal dimension (GFD) prescription of Hentschel and Proccacia \cite{hp} , Halsey et al.  \cite{hetal}. However, the presence of an even infinitesimal dissipation in the inertial range is shown to lead to a singularity in the GFD $D_q$ (at $q=1$) of the energy dissipation field and leads to a breakdown of the Meneveau-Sreenivasan \cite{ms9} binomial multiplicative formulation for a dissipative inertial cascade. The purpose of this paper is to demonstrate that this can be resolved by introducing a new appropriate \textit{ansatz} for the definition of the GFD $D_q$ to incorporate the effect of a scale-invariant dissipation via a phenomenological dissipative parameter $K ( 0 < K < 1)$.  This dissipation parameter is also shown to cause a steeper energy spectrum in the inertial range, as to be expected.  This ansatz is then generalized to incorporate a more symmetric dissipation via two dissipative parameters $K_1,$ and $K_2$ $( 0 < K_1$ and $K_2 < 1)$. 
\end{abstract}

\cleardoublepage
\section{Introduction}
Spatial intermittency effects in a fully developed turbulence (FDT) can be conveniently imagined to be related to the fractal aspects of the geometry of turbulence (Mandelbrot \cite{mandelbrot1}).  The mean energy dissipation field may then be assumed to be mimicked by a multi-fractal (Mandelbrot \cite{mandelbrot2}, Parisi and Frisch \cite{pf}). The underlying idea is that the energy dissipative structures are so strongly convoluted that they are not completely space filling (this idea received experimental support (Meneveau and Sreenivasan [4])). 

Multi-fractal models (Mandelbrot \cite{mandelbrot2}) generalize the notion of self-similarity from \textit{sets} to distributions of \tit{measures} defined on them. They are predicated on the idea that the singular measure (like the energy dissipation filed in FDT) resulting from a multiplicative fragmentation process exhibits a limiting \tit{scale-invariant} probability distribution. A multi-fractal comprises nested fractal subsets associated with different levels of this measure which are characterized by a continuous spectrum of scaling exponents (Hentschel and Proccacia \cite{hp}, Frisch and Parisi \cite{pf}, Halsey et al. \cite{hetal}; see Shivamoggi \cite{shivamoggi5}- \cite{shivamoggi8} for some fruitful applications of the multi-fractal models for various physical cases of FDT).

Meneveau and Sreenivasan \cite{ms9} gave a simple binomial model mimicking the underlying multi-fractal nature of the energy cascading process in the inertial range, and used the generalized fractal dimension (GFD) prescription of Hentschel and Proccacia \cite{hp} and Halsey et al. \cite{hetal}. This model fitted the entire spectrum of scaling exponents for the energy dissipation field in FDT amazingly well.  The one-dimensional (1D) version of this model was a generalized two-scale Cantor set with equal scales but unequal measures (with a ratio nearly equal to 7/3). This model was shown to exhibit all the measured multi-fractal properties of the 1D sections of the energy dissipation field. Meneveau and Sreenivasan \cite{ms9} mentioned the need for exploring further variant for their model which considers the possibility that eddies in the inertial range also dissipate some energy directly.  Such a possibility was also considered by Evertsz and Mandelbrot \cite{evertszMandelbrot}. However, the presence of an even infinitesimal dissipation in the inertial range is shown here to lead to a singularity in the GFD $D_q$ (at $q=1$) of the energy dissipation field and leads to a breakdown of the Meneveau-Sreenivasan \cite{ms9} binomial multiplicative formulation for a dissipative inertial cascade. The purpose of this paper (Shivamoggi et al. \cite{shivamoggi13}) is to trace this anomaly to the necessity to modify the GFD prescription of Hentschel and Proccacia \cite{hp} and Halsey et al. \cite{hetal} in dealing with a dissipative cascade and next to point out an appropriate modification of the GFD prescription to resolve this issue.

\section{Effect of Dissipation of Energy in the Inertial Range}
Let $E_{l}$ be the energy dissipation (or the energy flux out of) occurring in an eddy of size $l$. Following Meneveau and Sreenivasan \cite{ms9}, we stipulate a binomial multiplicative process to describe the energy flux distribution in the cascade of eddies in FDT.  We suppose that an eddy of size $l$ breaks down into $2^{d}$ ($d$ being the underlying space dimension, here we take it to be $1$) eddies of size $l/2$. We suppose that the energy flux to those small eddies is shared unequally in this binomial multiplicative process and partly dissipated locally at the same time, so a fraction $Kp$ ($K$ being the phenomenological dissipation parameter $0 < K < 1$ ) is distributed equally among half of the two new eddies and a fraction $(1-p)$ is distributed among the other half, so $ Kp+(1-p)<1 $.  This process is repeated over and over with fixed $K$ and $p$ over the inertial range. (A 1D section of this binomial multiplicative process is shown in Figure 3 of Reference \cite{ms9}.) It may be mentioned that $K$ is only a phenomenological representation of energy dissipation, and implies a scale-invariant process where a fixed amount of energy flux is extracted at all scales. This precludes correspondence to actual viscous dissipation which will be scale dependent and the multi-fractal in question will therefore not exhibit single inertial-range power-law scaling behavior.

Suppose $E_{l_n}$ is the energy dissipation that occurs, at the $nth$ stage of the cascade, in a box of size $l_n$, $l_n = L \left( 1/2 \right)^n$, where $L$ is the initial box size.  Assuming that the energy dissipation field is a multi-fractal, the GFD $D_q$ of the resulting energy flux distribution at the  nth stage of the energy cascade may be defined  following the prescription of Hentschel and Proccacia \cite{hp} and Halsey et al. \cite{hetal}. However, this leads to a singularity (see (\ref{4}) below) in the GFD $D_q$ (at $q=1$) in the presence of even an infinitesimal dissipation in the inertial range. In order to resolve this singularity we proceed to modify the prescription of Hentschel and Proccacia \cite{hp} and Halsey et al. \cite{hp} appropriately via a new \textit{ansatz} to incorporate the phenomenological dissipation parameter $K$, $(0<K<1),$ and hence render $D_q$ a well behaved function of $q$, $\forall$ q,

\begin{equation} \label{1}
\sum E_{l_{n}}^{q} = E_{L}^{q}\,\left(\frac{l_{n}}{L} \right)^{(\frac{q}{K}-1) D_q}.
\end{equation}

\noindent Here the coarse-grained probability measure given by the moments of the total energy dissipation occurring in a 3D box of size $\sc{l}_n$ is summed over $N(\sc{l}_n)$ boxes of size $\sc{l}_n$ contained in the support of this measure at the nth stage of the binomial multiplicative process. $E_L$ is the total energy dissipation in the initial box. 

This \tit{ansatz} is similar to Tsallis'\footnote{The non-extensive statistical mechanics approach to FDT is found to afford a whole new useful perspective to spatial intermittency aspects in FDT(Beck \cite{Beck}, Shivamoggi and Beck \cite{ShivamoggiBeck}, Shivamoggi \cite{Shivamoggi16}).} \cite{Tsallis} proposition to extend Boltzmann-Gibbs thermodynamics by considering entropy to be \tit{non-extensive} via a phenomenological non-extensivity parameter $q$. This prescription is suited for dealing with systems possessing long-range interactions.

At the nth stage of the cascade, the binomial sum of the $q$th moment of the energy dissipation over all eddies is (see \cite{ms9} and Appendix for more details),

\begin{equation} \label{2}
\sum E_{l_{n}}^{q} = E_{L}^{q} \, [ (Kp)^{q}+(1-p)^{q} ]^n.
\end{equation}

\noindent Combining (1) and (2), we obtain

\begin{equation} \label{3}
 E_{L}^{q} \, [ (Kp)^{q}+(1-p)^{q} ]^n = E_{L}^{q}\,(\frac{1}{2})^{n(\frac{q}{K}-1) D_q}
\end{equation}

\noindent from which, we obtain for the GFD $D_q$,

\begin{equation} \label{4}
D_q = \frac{ ln[(Kp)^{q} + (1-p)^{q} ] }{(1-\frac{q}{K}) \ln 2}.
\end{equation}
(\ref{4}) implies a singularity at $q/K = 1$, which may be precluded by restricting $q$ to integer values.

A full statistical characterization of the diversity of scaling in a multi-fractal requires an infinite spectrum of $D_q$ 's (Mandelbrot \cite{mandelbrot12}, Hentschel and Proccacia \cite{hp}).  Note that (4) gives for the fractal dimension $D_0$ of the \tit{support} of the energy dissipation measure, $ D_0 = 1 $, which of course implies that every new eddy that is produced in the cascade receives a non-zero energy flux even in a dissipative cascade.  Next, (4) gives for the \tit{entropy} dimension $D_1$,

\begin{equation} \label{5}
D_1 = \frac{ ln[ 1- p(1-K)] } {(1-\frac{1}{K}) \ln 2}
\end{equation}

The well-behaved expression for $D_1$,  given by (\ref{5}) validates the soundness of the new \textit{ansatz} for GFD $D_q$ given by (\ref{1}). This is further confirmed by the smoothness of the curve $D_q$ versus $q$ shown in Figure 1, for $p = 0.7$ and $ K = 0.8, 0.9, 1 $. It should be mentioned that $p = 0.7$ was found by Meneveau and  Sreenivasan \cite{ms9} to fit the experimental data very well.

\begin{figure}[H]
	\centerline{\includegraphics[width=6.0in]{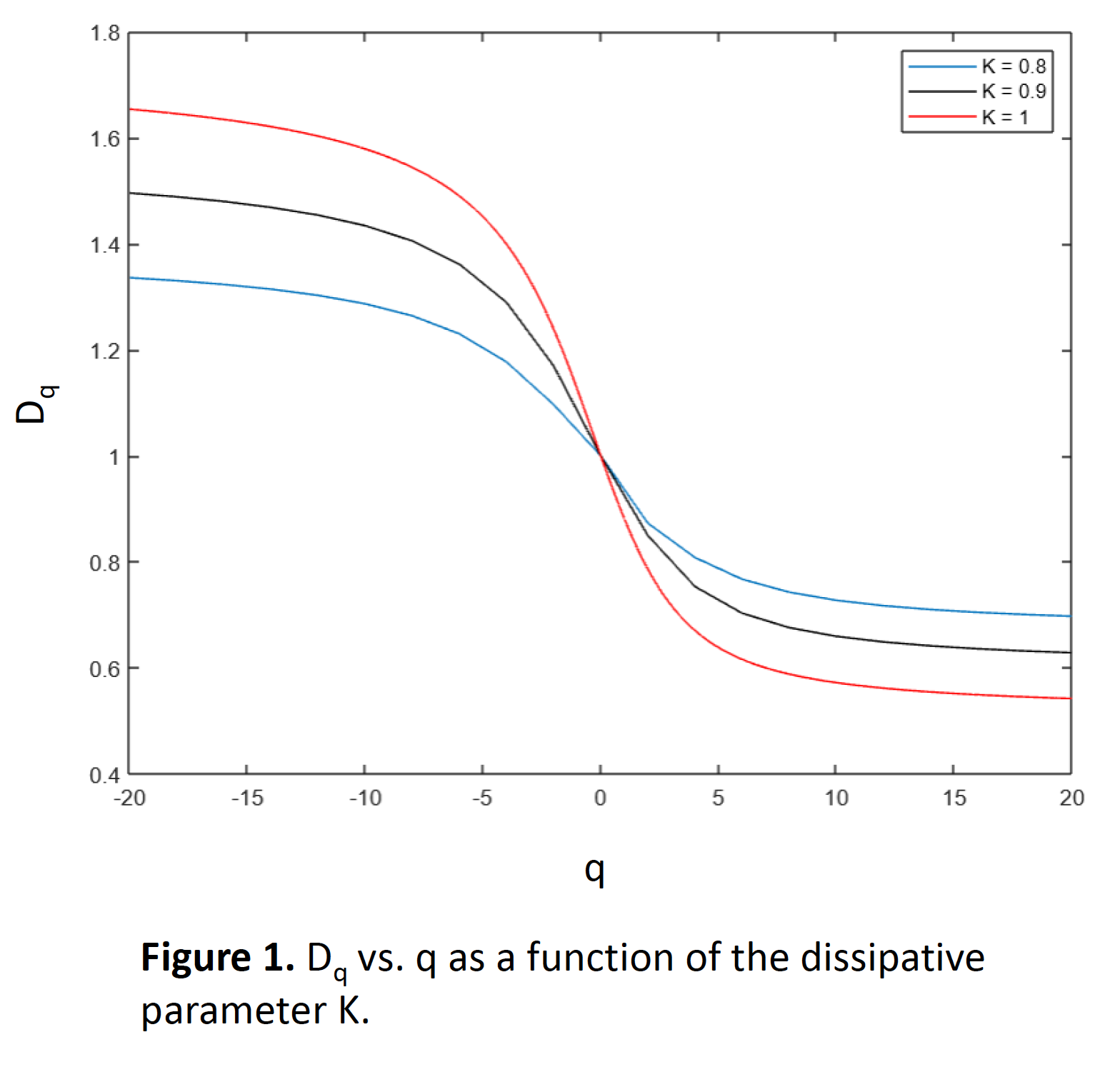}\\ }
	\label{fig:dsubq}
\end{figure}

\section{Energy Spectrum for the Dissipative Inertial Range}

We now provide a physical implication for the phenomenological dissipation parameter $K$ introduced in (1). For this purpose, we note that the singular measure in question, viz., the velocity increment over a distance $\sc{l}$ shows a scaling behavior,

\begin{equation}
\lim_{\sc{l} \to 0} |\delta v(\sc{l})| \sim \sc{l}^{\frac{\alpha}{3}}
\label{eq::6}
\end{equation}
and the singularities of order $\alpha/3$ are assumed to be \tit{uniformly} distributed on a fractal set $S(\alpha) \subset \mathbb{R}^3$ with Hausdorff dimension $f(\alpha)$. So, the number $N_{\sc{l}}(\alpha) d\mu(\alpha)$ of boxes of size $\sc{l}$ that correspond to values of $\alpha$ within a band $d \alpha$  around $\alpha$ scales as,

\begin{equation}
N_{\sc{l}} d \mu(\alpha) \sim \sc{l}^{-f(\alpha)} d\mu(\alpha)
\label{eq::7}
\end{equation}
where the measure $d \mu(\alpha)$ gives the weight of different scaling exponents of $\alpha$.

The velocity structure function $S_p(\sc{l})$, of order $p$, is obtained by space averaging $|\delta v(\sc{l})|^p$ over the entire volume. Noting that the total number of boxes of size $\sc{l}$ in $\mathbb{R}^3$ is proportional to $\sc{l}^{-3}$, the probability to belong to the set $S(\alpha)$ at scale $\sc{l}$ is proportional to $\sc{l}^{3-f(\alpha)}$. Thus,

\begin{equation}
S_p(\sc{l}) \equiv \langle |\delta v(\sc{l})|^p \rangle \sim \int d \mu(\alpha) \sc{l}^{3-f(\alpha)+\frac{p \alpha}{3}} \sim \sc{l}^{\zeta_p}
\label{eq::8}
\end{equation}
where $\zeta_p$ is the characteristic exponent of the velocity structure function of order $p$.

In the limit $\sc{l} \to 0$, the contribution from the smallest exponent dominates the integral in $(8)$, so on using the method of steepest descent to extract the dominant term in the latter, we obtain,

\begin{equation}
\zeta_p = \inf_{\alpha} \left[ 3-f(\alpha)+\frac{\alpha p}{3} \right] = 3-f(\alpha^*)+ \frac{\alpha^* p}{3}
\label{eq::9}
\end{equation}
where,

\begin{equation}
\frac{d f(\alpha^*)}{d \alpha} = \frac{p}{3}.
\label{eq::10}
\end{equation}

On the other hand, upon approximating the sum on the left hand side in $(1)$ by an integral over all possible values of $\alpha$, we obtain

\begin{equation}
\int d \mu(\alpha) \sc{l}^{(\alpha+2) q-f(\alpha)} \sim \sc{l}^{\left( \frac{q}{K}-1 \right) D_q}, \hspace{.1in} \sc{l} \hspace{.1in} \text{small.}
\label{eq::11}
\end{equation}
On extracting the dominant term in the integral above, in the limit $\sc{l} \to 0$, via the method of steepest descent, we obtain

\begin{equation}
(\hat{\alpha}+2)q-f(\hat{\alpha}) = \left( \frac{q}{K}-1 \right) D_q
\label{eq::12}
\end{equation}
where,

\begin{equation}
\frac{d f(\hat{\alpha})}{d \alpha} = q.
\label{eq::13}
\end{equation}

We now assume that the values of $\alpha^*$ and $\hat{\alpha}$ given by (\ref{eq::10}) and (\ref{eq::13}), for which the integrands in (\ref{eq::8}) and (\ref{eq::11}) become extremum, are coincident, as per the \textit{Kolmogorov refined similarity hypothesis} (Kolmogorov \cite{kolmogorov}, Meneveau and Sreenivasan \cite{ms4}).

Eliminating $f(\alpha)$ from $(9)-(13)$, we obtain

\begin{equation}
\zeta_p = \frac{p}{3}-\frac{1}{3}(p-3)(3-D_{p/3})+\frac{p}{3} \left(\frac{1}{K}-1 \right)D_{p/3}
\label{eq::14}
\end{equation}
from which, the energy spectrum (corresponding to $p=2$) is given by

\begin{equation}
E(k) \sim k^{-\frac{5}{3}-\frac{1}{3}(3-D_{2/3})-\frac{2}{3} \left(\frac{1}{K}-1 \right)D_{2/3}}, \hspace{.1in} 0<K<1.
\label{eq::15}
\end{equation}
$(15)$ shows that the energy spectrum is steeper for a dissipative inertial cascade $(0<K<1)$, as to be expected, and indicated also by the shell model numerical calculation (Leveque and She \cite{ls}).

\section{Generalized Treatment of Dissipation of Energy in the Inertial Range}

Let us now relax the asymmetry in introducing energy dissipation in the inertial range via a single phenomenological dissipation parameter $K$ in Section 2, and consider the more general case. Let us suppose that the energy flux to the new eddies of size $l/2$ is shared equally and partly dissipated locally at the same time, so a fraction $K_1p$ $(K_1 < 1)$ is distributed equally among half of the new eddies and a fraction $K_2 (1-p)$ is distributed equally among the other half so $K_1p+K_2(1-p) < 1$ ($K_1$ and $K_2$ being two phenomenological dissipative parameters). This process is repeated over and over again with fixed $K_1, K_2$ and $p$ over the inertial range. (2) then becomes,

\begin{equation} \label{6}
\sum E_{l_{n}}^{q} = E_{L}^{q} \, [ (K_1 p)^{q}+K_2^{q}(1-p)^{q} ]^n.
\end{equation}

We proceed now to define for this general case, the generalized fractal dimension $D_q$ of the resulting energy flux distribution at the nth stage of the energy cascade by incorporating two phenomenological dissipative parameters $K_1$ and $K_2$ (in place of (1) ) as,

\begin{equation} \label{7}
\sum E_{l_{n}}^{q} = E_{L}^{q}\,(\frac{l_{n}}{L})^{(\frac{q}{K_1/K_2}-1) D_q}
\end{equation}

\noindent Combining (16) and (17) we obtain

\begin{equation} \label{8}
E_{L}^{q} \, [ (K_1 p)^{q}+K_{2}^{q}(1-p)^{q} ]^n = E_{L}^{q}\,(\frac{1}{2})^{n(\frac{q}{K_1/K_2}-1) D_q}
\end{equation}

\noindent from which,

\begin{equation} \label{9}
D_q = \frac{ ln[(K_1 p)^{q} + K_{2}^{q}(1-p)^{q} ] }{(1-\frac{q}{K_1/K_2}) ln2}
\end{equation}

\noindent Figure 2 shows $D_q$ vs. q for $p = 0.7, K_1 = 0.8$  and $K_2 = 0.9$. The smoothness of the curve $D_q$ versus $q$ implying that $D_q$ is a well-behaved function of $q$ validates the soundness of the new \textit{ansatz} for GFD $D_q$ given by (\ref{7}) for the generalized dissipation case.

\section{Discussion}
The nature of viscous effects on the inertial range dynamics is not well understood while statistical states in the inertial range are believed to depend on the viscous dissipation mechanism (Leveque and She \cite{ls}). In this paper, we explore a variant for the simple model based on a binomial multiplicative process of Meneveau and Sreenivasan \cite{ms9} that mimics the multi-fractal nature of the energy dissipation field in the inertial range of fully developed turbulence (FDT). This involves including the possibility that eddies in the inertial range also dissipate some energy directly.  However, the presence of an even infinitesimal dissipation in the inertial range is shown to lead to a singularity in the GFD $D_q$ (at $q=1$) of the energy dissipation field and hence a breakdown of the Meneveau-Sreenivasan \cite{ms9} binomial multiplicative formulation for a dissipative inertial cascade. This is shown to be resolvable by introducing a new appropriate \textit{ansatz}, given by (\ref{1}) and (\ref{eq::7}) for the definition of the generalized fractal dimension (GFD) $D_q$ to incorporate the effect of a scale-invariant dissipation via a phenomenological dissipative parameter $K ( 0 < K < 1)$.  This phenomenological dissipation parameter is also shown to cause a steeper energy spectrum in the inertial range, as to be expected. This \textit{ansatz}, given by (\ref{1}) and (\ref{eq::7}), is then generalized to incorporate a more symmetric dissipation via two phenomenological dissipative parameters $K_1$ and $K_2$ $(0 < K_1$ and $K_2 < 1)$.  The soundness of the new GFD \tit{ansatz}, given by (\ref{1}) and (\ref{eq::7}), is validated by the resolution of the singularities in the GFD $D_q$ (at $q=1$) in the Meneveau-Sreenivasan \cite{ms9} binomial multiplicative formulation as well as the steepening of the energy spectrum caused by the dissipation in the inertial range. 

It may be noted, as Stanley and Meakin \cite{sm} pointed out, that the left hand side in (\ref{1}) is formally analogous to the \textit{partition function} $Z(\beta)$ in thermodynamics so that $q/K$ is like $\beta \sim 1/T$, $T$ being the \textit{temperature}, and $D_q$ is like the \textit{free energy}. The \textit{Legendre transform} $f(\alpha)$ is the analogue of the \textit{entropy} $S$, with $\alpha$ being the analogue of the \textit{internal energy}. So, (\ref{1}) implies a lower effective \textit{temperature} compared to that for the non-dissipative case. On the other hand, the presence of dissipative effects in the present model would imply a lower level of singularities in the nonlinear dynamics underlying the FDT case, and hence a lower \tit{entropy} (Chabbra et al.  \cite{cj}) in the thermodynamic system formally equivalent to the present FDT system. Thus, the dissipation parameter lowers the singularity spectrum $f(\alpha)$, which implies that the effective \tit{entropy} decreases as the effective \tit{temperature} is lowered for a multi-fractal model. This was confirmed by T\'el \cite{tel}.

It may be mentioned that in view of the uncertainties involved and lack of clear and reliable guidelines in choosing appropriate values for the phenomenological parameters in the present model, it does not seem feasible at this point to do quantitative comparison of the present results with experimental or numerical data on grid, wake and atmospheric turbulence.

\setcounter{figure}{1}

\begin{figure}[H]
	\centerline{\includegraphics[width = 6.0 in]{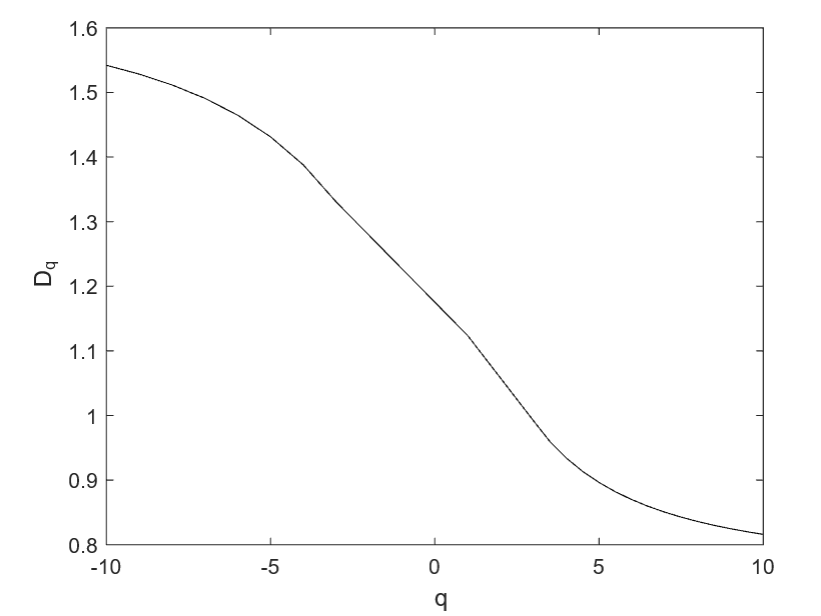}\\ }
	\caption{\large $D_q$ vs. $q$ as a function of the dissipative parameters $K_1 = 0.8, K_2 = 0.9$.}
	\label{fig:K1K2}
\end{figure}

\section*{Appendix: The Binomial Multiplicative Process}
The binomial multiplicative process is a purely discrete process associated with an experiment involving two possible outcomes (Mazo \cite{mazo}) - say, an event $A$ and an event $B$, with respective probabilities $p$ and $(1-p)$ in each experiment. Suppose the experiment is performed $n$ times with each trial being independent of all other trials. Noting that the events $A$ and $B$ can occur in any order, we have that the probability for a sequence with \textit{exactly} $m$ occurrences of event $A$ is $p^m(1-p)^{n-m}$ and there are $\binom{n}{m}=\frac{n!}{m!(n-m)!}$ such sequences for a given $m$. The probability of \textit{exactly} $m$ occurrences of event A out of $n$ independent trials is then given by
$$
\binom{n}{m}p^m(1-p)^{n-m}
$$
which is the \textit{binomial distribution} with \textit{mean} $np$ and \textit{variance} $np(1-p)$.

Coming back to the discussion in Section 2, at the $nth$ step of the energy cascade, there will therefore be $\binom{n}{m}$ eddies of size $\sc{l}_n = L\left( \frac{1}{2} \right)^n$ and the $q$th moment of the energy dissipation occurring in each eddy is given by,
$$
E_{\sc{l}_n}^q = (Kp)^{mq}(1-p)^{(n-m)q}E_L^q, \hspace{.2in} 0<m<n.
$$
The $q$th moment of the total energy dissipation at the $nth$ stage is then given by summing $E_{\sc{l}_n}^q$ over all the $\binom{n}{m}$ eddies, which is just the \textit{binomial sum} given by (\ref{2}).

\section*{Acknowledgements}
BKS expresses his sincere thanks to Professors K.R. Sreenivasan and Charles Meneveau for their valuable suggestions and advice, and the referees for their helpful remarks.

\end{document}